\documentclass[final,3p,times,onecolumn]{elsarticle}

\usepackage{amssymb}

\usepackage{amsmath}
\usepackage{setspace}
\journal{Journal of Nuclear Instruments and Methods in Physics Research A}
\usepackage{xcolor}
\usepackage{url}

\begin{document}

\begin{frontmatter}



\title{Evaluation of the performance of event reconstruction algorithms in the JSNS$^2$ experiment using a $^{252}$Cf calibration source}


\author[a]{D. H. Lee}
\author[b]{M. K. Cheoun}
\author[c]{J. H. Choi}
\author[d]{J. Y. Choi}
\author[e,h]{T. Dodo}
\author[f]{J. Goh}
\author[g]{K. Haga}
\author[g]{M. Harada}
\author[h,g]{S. Hasegawa}
\author[f]{W. Hwang}
\author[i]{T. Iida}
\author[d]{H. I. Jang}
\author[j]{J. S. Jang}
\author[k]{K. K. Joo}
\author[l]{D. E. Jung}
\author[m]{S. K. Kang}
\author[g]{Y. Kasugai}
\author[n]{T. Kawasaki}
\author[o]{E. J. Kim}
\author[k]{J. Y. Kim}
\author[p]{S. B. Kim}
\author[q]{W. Kim}
\author[g]{H. Kinoshita}
\author[n]{T. Konno}
\author[k]{I. T. Lim}
\author[r]{C. Little}
\author[a]{T. Maruyama} 
\author[r]{E. Marzec}
\author[g]{S. Masuda}
\author[g]{S. Meigo}
\author[k]{D. H. Moon}
\author[s]{T. Nakano}
\author[t]{M. Niiyama}
\author[a]{K. Nishikawa}
\author[c]{M. Y. Pac}
\author[k]{H. W. Park}
\author[q]{J. S. Park}
\author[m]{R. G. Park}
\author[u]{S. J. M. Peeters}
\author[v,l]{C. Rott}
\author[g]{K. Sakai}
\author[g]{S. Sakamoto}
\author[s]{T. Shima}
\author[a]{C. D. Shin}
\author[r]{J. Spitz}
\author[e]{F. Suekane}
\author[s]{Y. Sugaya}
\author[g]{K. Suzuya}
\author[i]{Y. Takeuchi}
\author[g]{Y. Yamaguchi}
\author[w]{M. Yeh}
\author[c]{I. S. Yeo}
\author[f]{C. Yoo}
\author[l]{and I. Yu}

\affiliation[a]{organization={High Energy Accelerator Research Organization (KEK)},
            addressline={1-1 Oho}, 
            city={Tsukuba},
            postcode={305-0801}, 
            state={Ibaraki},
            country={Japan}}
\affiliation[b]{organization={Department of Physics, Soongsil University},
            addressline={369 Sangdo-ro}, 
            city={Dongjak-gu},
            postcode={06978}, 
            state={Seoul},
            country={Korea}}
\affiliation[c]{organization={Laboratory for High Energy Physics, Dongshin University},
            addressline={67, Dongshindae-gil}, 
            city={Naju-si},
            postcode={58245}, 
            state={Jeollanam-do},
            country={Korea}}
\affiliation[d]{organization={Department of Fire Safety, Seoyeong University},
            addressline={1 Seogang-ro}, 
            city={1 Seogang-ro},
            postcode={61268}, 
            state={Gwangju},
            country={Korea}}
\affiliation[e]{organization={Research Center for Neutrino Science, Tohoku University},
            addressline={6-3 Azaaoba, Aramaki}, 
            city={Sendai},
            postcode={980-8578}, 
            state={Aoba-ku},
            country={Japan}}
\affiliation[f]{organization={Department of Physics, Kyung Hee University},
            addressline={26, Kyungheedae-ro}, 
            city={Seoul},
            postcode={02447}, 
            state={Dongdaemun-gu},
            country={Korea}}
\affiliation[g]{organization={J-PARC Center, JAEA},
            addressline={2-4 Shirakata, Tokai-mura}, 
            city={Naka-gun},
            postcode={319-1195}, 
            state={Ibaraki},
            country={Japan}}
 \affiliation[h]{organization={Advanced Science Research Center, JAEA},
            addressline={2-4 Shirakata, Tokai-mura}, 
            city={Naka-gun},
            postcode={319-1195}, 
            state={Ibaraki},
            country={Japan}}           
 \affiliation[i]{organization={Faculty of Pure and Applied Sciences, University of Tsukuba},
            addressline={ Tennodai 1-1-1}, 
            city={Tskuba},
            postcode={305-8571}, 
            state={Ibaraki},
            country={Japan}}  
 \affiliation[j]{organization={Department of Physics and Photon Science, Gwangju Institute of Science and Technology},
            addressline={123 Cheomdangwagi-ro}, 
            city={Buk-gu},
            postcode={61005}, 
            state={Gwangju},
            country={Korea}}
 \affiliation[k]{organization={Department of Physics, Chonnam National University},
            addressline={77, Yongbong-ro}, 
            city={Buk-gu},
            postcode={61186}, 
            state={Gwangju},
            country={Korea}}           
 \affiliation[l]{organization={Department of Physics, Sungkyunkwan University},
            addressline={2066, Seobu-ro, Jangan-gu}, 
            city={Suwon-si},
            postcode={16419}, 
            state={Gyeonggi-do},
            country={Korea}}  
 \affiliation[m]{organization={School of Liberal Arts, Seoul National University of Science and Technology},
            addressline={232 Gongneung-ro}, 
            city={Nowon-gu},
            postcode={139-734}, 
            state={Seoul},
            country={Korea}}    
 \affiliation[n]{organization={Department of Physics, Kitasato University},
            addressline={1 Chome-15-1 Kitazato Minami Ward}, 
            city={Sagamihara},
            postcode={252-0329}, 
            state={Kanagawa},
            country={Japan}} 
 \affiliation[o]{organization={Division of Science Education, Jeonbuk National University},
            addressline={567 Baekje-daero, Deokjin-gu}, 
            city={Jeonju-si},
            postcode={54896}, 
            state={Jeollabuk-do},
            country={Korea}} 
 \affiliation[p]{organization={School of Physics, Sun Yat-sen (Zhongshan) University},
            addressline={Haizhu District}, 
            city={Guangzhou},
            postcode={510275}, 
            state={},
            country={China}} 
 \affiliation[q]{organization={Department of Physics, Kyungpook National University},
            addressline={80 Daehak-ro}, 
            city={Buk-gu},
            postcode={41566}, 
            state={Daegu},
            country={Korea}} 
 \affiliation[r]{organization={University of Michigan},
            addressline={500 S. State Street}, 
            city={Ann Arbor},
            postcode={48109}, 
            state={MI},
            country={U.S.A}} 
  \affiliation[s]{organization={Research Center for Nuclear Physics, Osaka University},
            addressline={10-1 Mihogaoka}, 
            city={Ibaraki},
            postcode={567-0047}, 
            state={Osaka},
            country={Japan}} 
  \affiliation[t]{organization={Department of Physics, Kyoto Sangyo University},
            addressline={Motoyama, Kamigamo}, 
            city={Kita-ku},
            postcode={603-8555}, 
            state={Kyoto},
            country={Japan}} 
 \affiliation[u]{organization={Department of Physics and Astronomy, University of Sussex},
            addressline={Falmer}, 
            city={Brighton},
            postcode={BN1 9RH}, 
            state={East Sussex},
            country={UK}}
 \affiliation[v]{organization={Department of Physics and Astronomy, University of Utah},
            addressline={201 Presidents' Cir}, 
            city={Salt Lake City},
            postcode={84112}, 
            state={UT},
            country={U.S.A}}
 \affiliation[w]{organization={Brookhaven National Laboratory},
            addressline={Upton}, 
            city={},
            postcode={11973-5000}, 
            state={NY},
            country={U.S.A}}
            
\begin{abstract}

JSNS$^2$ investigates short-baseline neutrino oscillations using a 24-meter baseline and a 17-tonne Gd-loaded liquid scintillator target. Accurate event-reconstruction algorithms are crucial for analyzing experimental data. The algorithms undergo meticulous validation through calibration with a $^{252}$Cf source. This paper outlines the methodology and evaluates the reconstruction performance, focusing on neutrino interactions up to approximately 50~MeV for sterile neutrino searches. Both $^{252}$Cf and Michel electron events are studied to evaluate reconstruction accuracy. The analysis concludes that the uncertainty of the fiducial volume, with an appropriate correction, is much less than the requirement of JSNS$^2$ requirement (10\%). Furthermore, the energy resolution is measured to be 3.3$\pm$0.1\% for the Michel electron endpoint and 4.3$\pm$0.1\% for the n-Gd peak in the central region.

\end{abstract}




\end{frontmatter}


\section{Introduction}
\label{sec:introduction}
The potential existence of sterile neutrinos has been a seminal topic of investigation in neutrino physics for more than two decades.
The experimental results of \cite{CITE:LSND, CITE:BEST, CITE:MiniBooNE2018, CITE:REACTOR} could be interpreted as indications of sterile neutrinos with mass-square differences on the order of $1\ \mathrm{eV}^2$.

The JSNS$^2$ experiment, which was proposed in 2013~\cite{CITE:JSNS2proposal}, 
searches for neutrino oscillations with a short baseline caused by sterile neutrinos at the Material and Life science experimental Facility (MLF~\cite{CITE:MLF}) in J-PARC.
The facility provides an intense and high-purity source of neutrinos generated via muon decay-at-rest ($\mu$DAR). 
These neutrinos are produced by the collision of 1~MW 3~GeV
protons from a rapid cycling synchrotron onto a mercury target with a 25~Hz repetition rate within the MLF.
The short-pulsed beam, with a low duty factor (0.6~$\mu$s beam pulse / 40,000~$\mu$s pulse separation), enables excellent background rejection. Only neutrinos from muon decay-at-rest persist beyond 1~$\mu$s from the beam start, effectively eliminating those from pion and kaon decay-in-flight. 
Furthermore, surviving negative pions are absorbed through pionic atom formation, and negative muons are captured by nuclei into muonic atoms. Based on Geant4 simulation~\cite{CITE:Geant4} around the mercury target, $\sim$99\% of negative pions are immediately absorbed into atoms. In addition to this, negative muon captures happen for $\sim$80\% of total. This results in a substantial suppression of neutrinos from both negative pions and negative muons, in the order of $10^{-3}$~\cite{CITE:TDR}.  

JSNS$^2$ employs a cylindrical liquid scintillator detector with a diameter of 4.6~m and a height of 3.5~m, located at a distance of 24~m from the mercury target in the MLF. The detector consists of 17~tonnes of Gadolinium (Gd) loaded liquid scintillator (Gd-LS) detector with 0.1\% Gd concentration contained within an acrylic vessel, and 33~tonnes unloaded liquid scintillator (LS) located between the acrylic vessel and a stainless steel tank. Both LS and Gd-LS use Linear Alkyl Benzene (LAB) as the base solvent, 3~g/L PPO (2,5-diphenyloxazole) as the fluor, and 15~mg/L 1,4-bis(2-methylstyryl) benzene (bis-MSB) as the wavelength shifter. An optical separator divides the LS volume into two independent layers, forming two detector volumes within a single detector.

The region inside the optical separator, known as the `inner detector', includes the entire volume of the Gd-LS and an approximately 25~cm thick LS layer. Scintillation light from the inner detector is detected by 96 Hamamatsu R7081 photomultiplier tubes (PMTs), each with a 10-inch diameter.

The outer layer, known as the `veto layer', is designed to detect cosmic-ray induced particles crossing the detector. This layer is equipped with 24 PMTs, each 10 inches in diameter. Twelve PMTs are positioned in the top region, and the remaining 12 are in the bottom veto region. 

Further details of the JSNS$^2$ detector can be found in \cite{CITE:NIM}.

The energy and vertex reconstruction of JSNS$^2$ is performed using
JADE (JSNS$^2$ Analysis Development Environment), as detailed in~\cite{CITE:JADE}. JADE employs the maximum likelihood technique
and serves as the basis for the JSNS$^2$ data analyses. It also determines the fiducial volume for physics results, making its verification essential for the experiment. The precision requirement for the fiducial volume is 10\% for cross-section measurements.

\section{$^{252}$Cf Source Calibration}
\label{sec:method}
A calibration using neutrons from a $^{252}$Cf radioactive source was 
performed to evaluate the JADE performance at various positions within the detector. The $^{252}$Cf source, with dimensions of 
20~mm in length and 5~mm in diameter, had a calibrated activity of 3.589 $\times$ $10^{6}$ Bq on the 10\textsuperscript{th} of August 1983. Neutrons are emitted during the source's spontaneous fission, constituting about 3.1\% of this total activity. 

The dominant neutron capture is expected to be on $^{157}$Gd, which has a natural abundance of 15.65\% and a capture cross-section of 254,000~b, resulting in a total gamma energy of 7.937~MeV. A weaker second peak is expected at 8.536~MeV from the capture on $^{155}$Gd, which has a natural abundance of 14.80\% and a capture cross-section of 60,900~b~\cite{CITE:CF}. Based on these abundances and cross sections, approximately 80\% of captures occur on $^{157}$Gd, while 20\% occur on $^{155}$Gd.

Calibrations with the $^{252}$Cf source are also crucial for understanding the delayed signal in Inverse Beta Decay (IBD) event searches. In IBD, neutrons serve as the delayed signal and are captured on Gd, paralleling the capture process in the $^{252}$Cf calibration.

\begin{figure}[h]
\begin{center}
\includegraphics[width=0.40\textwidth]{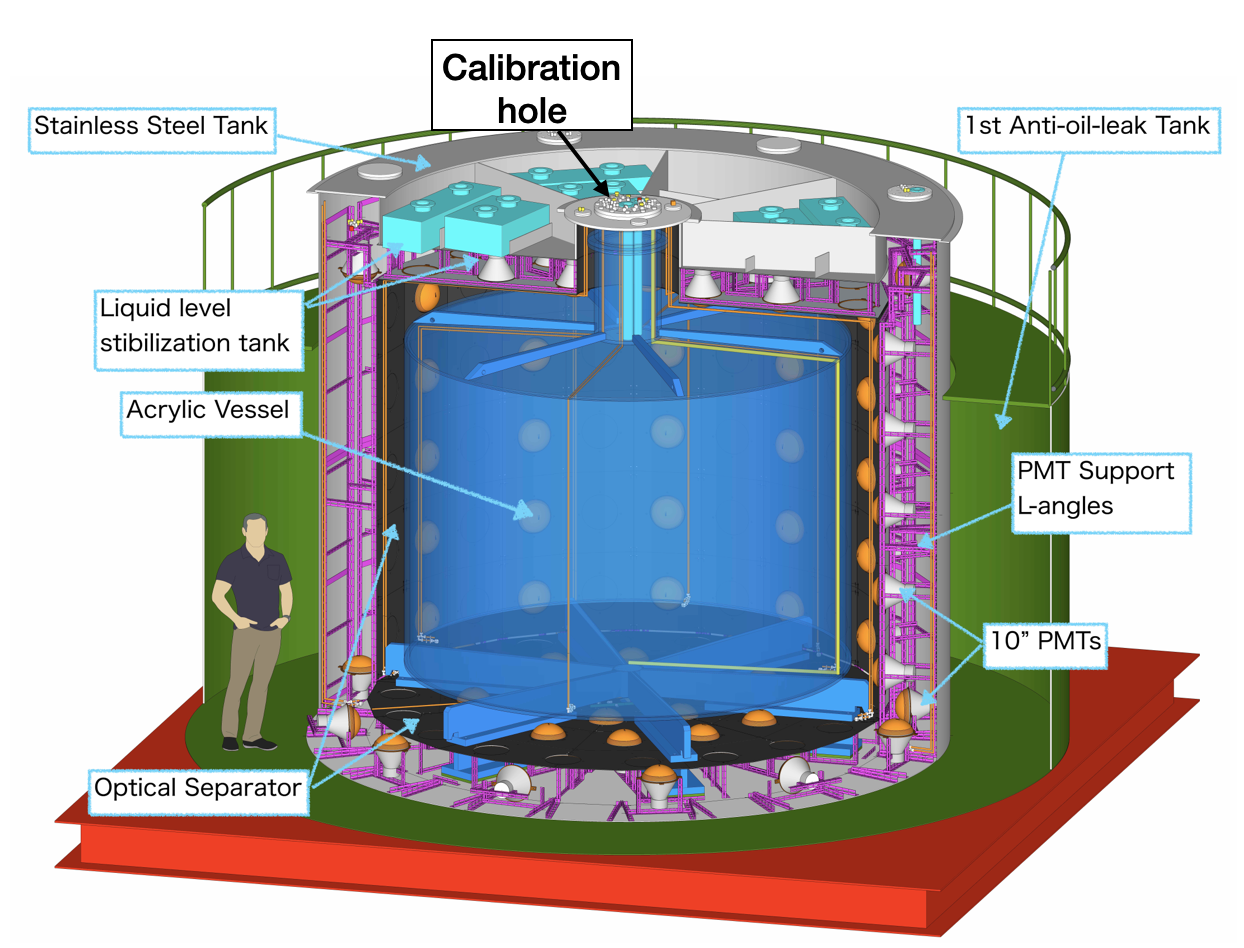} 
\caption{\label{fig:calib_hole} A schematic of the JSNS$^2$ detector, including the calibration access hole.} 
\end{center}
\end{figure}

\begin{figure}[h]
\begin{center}
\includegraphics[width=0.40\textwidth]{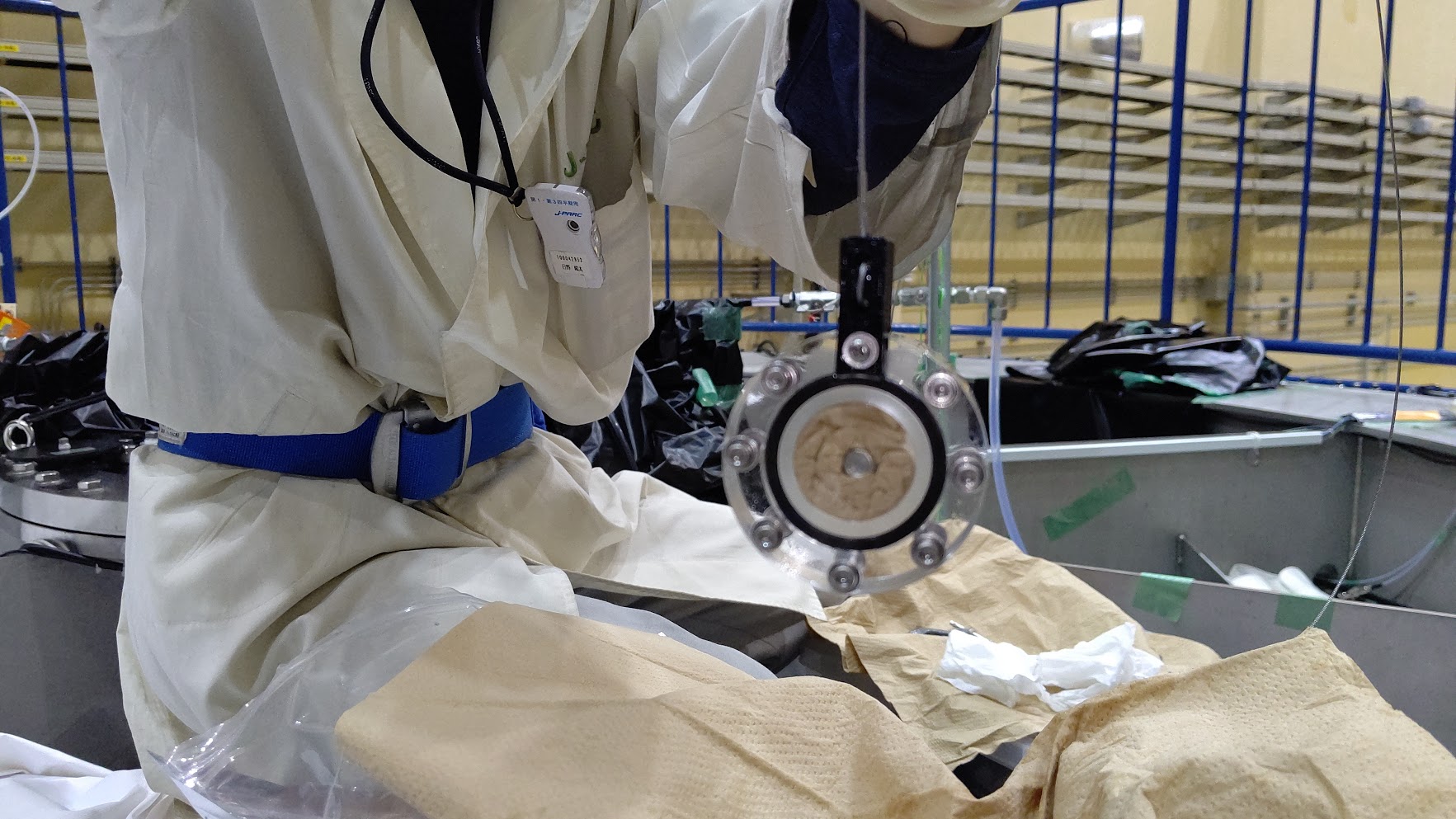} 
\caption{\label{fig:acrylic_container} Acrylic container for the one-dimensional calibration system.} 
\end{center}
\end{figure}

One- and three-dimensional $^{252}$Cf calibration systems have been developed for JSNS$^2$. A crucial aspect of this calibration is utilizing the 70~mm diameter calibration access hole in the detector, as illustrated in Figure~\ref{fig:calib_hole}. 
For the one-dimensional calibration system, the neutron source is housed in an acrylic cylindrical holder, as shown in Figure~\ref{fig:acrylic_container}. The holder has a diameter of 60~mm and a thickness of 16~mm. 
When housed in the acrylic container, the neutron source can be positioned at any location along the detector's central axis using a stepping motor. 
The one-dimensional calibration system is described in~\cite{CITE:NIM}.

\begin{figure}[h]
\begin{center}
\includegraphics[width=0.40\textwidth]{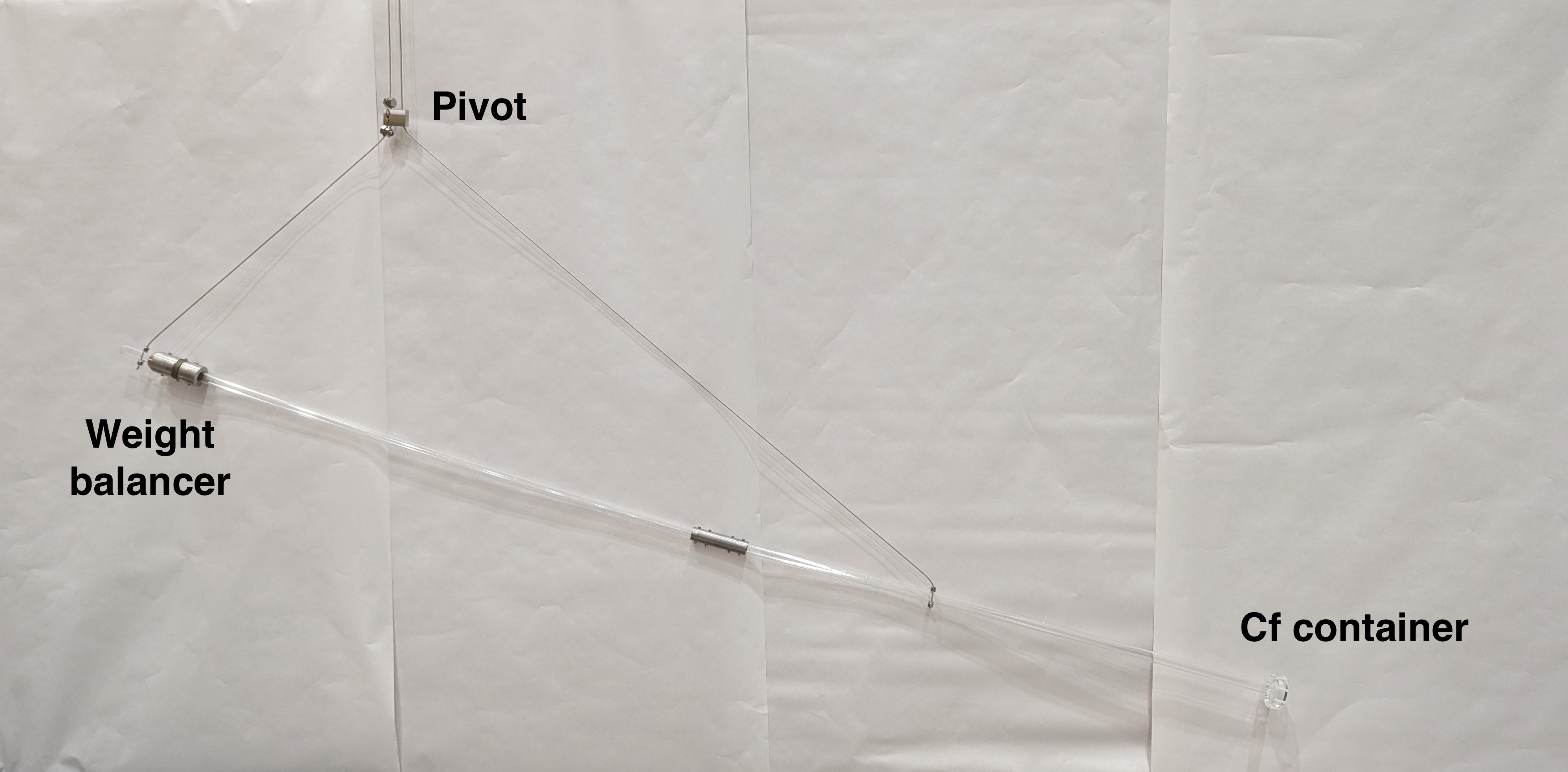} 
\caption{\label{fig:3D} The three-dimensional calibration system.} 
\end{center}
\end{figure}

The three-dimensional system is newly constructed and consists of an acrylic pipe, a container, a stainless balance weight, wires, and a wire pivot, as shown in Figure~\ref{fig:3D}. The acrylic container, chemically bonded to the pipe, is the same size as the one-dimensional calibration system. The acrylic pipe has a diameter of 20~mm with a thickness of 2~mm and a length of 1000~mm and 2200~mm
for calibrations at a radius $R$ = 40~cm and 160~cm, respectively. Here, $R$ is defined as $R = \sqrt{x^2 + y^2}$, with the origin of the coordinate system at the detector's center. 

The lengths of the pipes can be adjusted to change $R$ of the calibration points. The wires are secured by stainless steel bolts and nuts, positioned at 30~cm (or 60~cm) from the $^{252}$Cf container side and 
2.0~cm (or 4.0~cm) from the opposite side. The values in parentheses are for the $R$ = 160~cm calibration, while the other values apply to the $R$ = 40~cm calibration.

A weight balance is used for the $R$ = 160~cm calibration to keep the acrylic pipe as horizontal as possible. Without the weight balance, the $^{252}$Cf container side cannot be elevated. The weight balance, attached to the opposite side, weighs 760~g and is made of passivated SUS304 stainless steel. It is fixed to the acrylic pipe with stainless bolts and nuts. Note that the balance is not required for the $R = 40$~cm calibration.

The pivot has two 4~mm diameter holes for the wires to pass through and is also made of stainless steel. The compatibility of passivated stainless steel with Gd-LS was demonstrated in \cite{CITE:Hino}.

\begin{figure}[h]
\begin{center}
\includegraphics[width=0.45\textwidth]{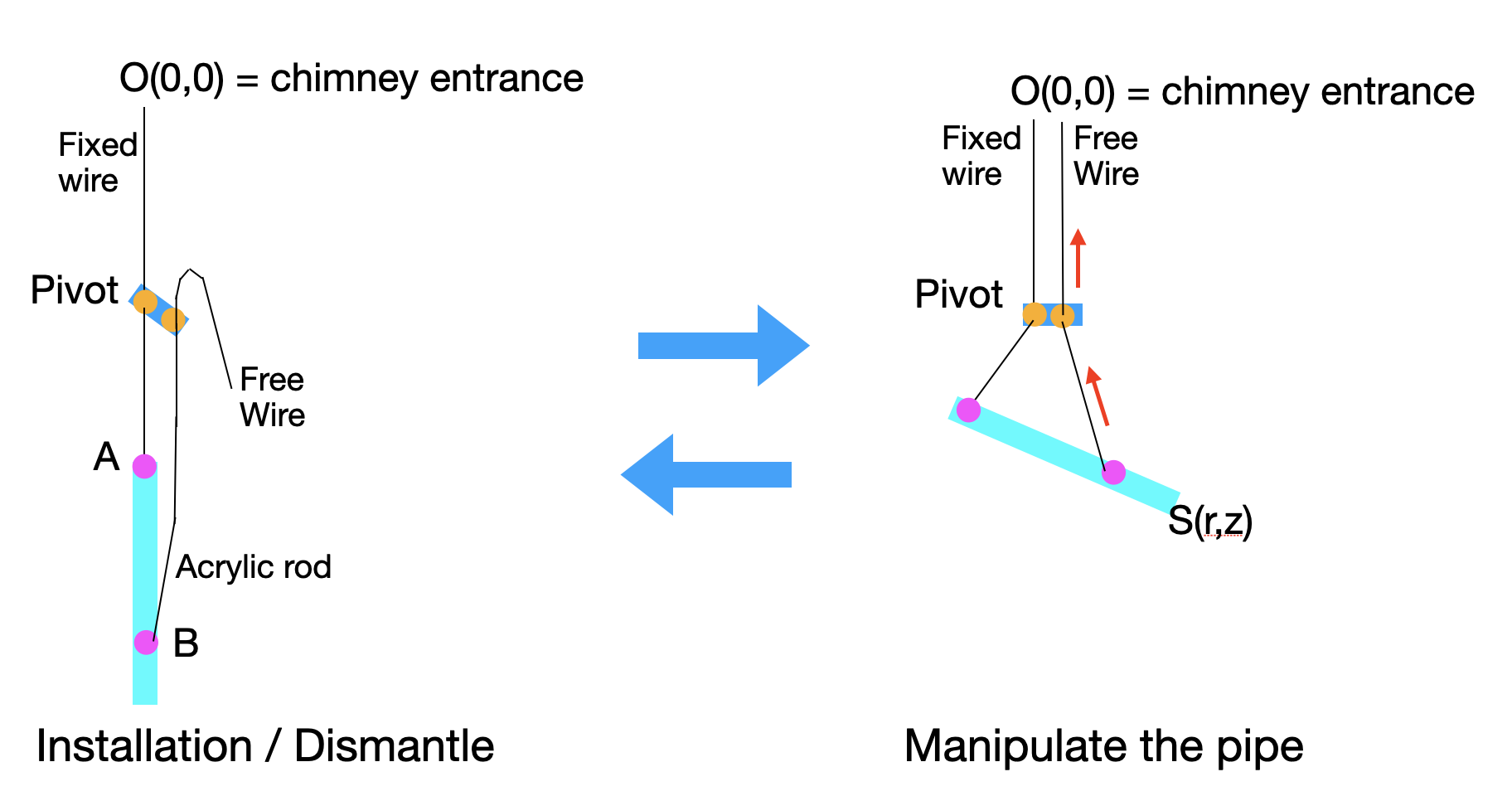} 
\caption{\label{fig:3Dprinciple} This sketch illustrates the principle of the three-dimensional calibration system. During installation, the pipe is initially aligned vertically within the detector. Once inside, the angle of the acrylic pipe can be adjusted to position the $^{252}$Cf source at any desired location. Upon dismantling the system, the pipe is returned to its vertical alignment.} 
\end{center}
\end{figure}
Figure~\ref{fig:3Dprinciple} shows the principle of the three-dimensional calibration system, which is inspired by KamLAND’s system~\cite{CITE:KAMLAND}. During installation, the free wire is untensioned and is used to manually change the angle of the acrylic pipe after immersing the pivot into the Gd-LS. When fully extended, the pipe is horizontally aligned within the detector, allowing the $^{252}$Cf~source to be positioned anywhere in the target volume (R $<$ 1.6~m and $|Z| <$ 1.25~m). The time required for the three-dimensional calibration is minimized to reduce the risk of oxygen contamination. 

For this study, calibration data was collected at $|z|$ positions of 0, 25, 50, 75, and 100~cm for the one-dimensional calibration, and ($R,z$) = (40~cm, -7.0~cm) and (160.0~cm, 9.0~cm) for the three-dimensional system.
The precision of the calibration points is 1~cm. For the one-dimensional system, relative source positions determined by the stepping motor's wire length depend on the precision of the production of acrylic and stainless-steel tank production.
To calibrate the true source positions of the three-dimensional system, it was manipulated in the air outside the detector. Multiple measurements of the source positions were performed using a tape measure ruler with a precision of 1~mm. However, beyond this precision, reproducibility becomes a primary factor in the systematic uncertainties, achieving a positional accuracy of 1 cm for the 1000~mm acrylic pipe and 2 cm for the 2200~mm acrylic pipe.

For the 2200~mm acrylic pipe, the measurement of source position in the air is $R$ = 150~cm. However, in the liquid scintillator (LS), the buoyant effect acting on the system is not uniformly distributed, as the weight balance - made of stainless steel - is located on only one side. This unbalanced buoyant effect causes the source positions to differ between the air and LS. Consequently, the source position is assumed to be located at $R$ = 160~cm, with 1~cm precision based on the acrylic vessel’s production constraints. For the 1000~mm acrylic pipe, the source position remains fixed at $R$ = 40~cm, with a precision of 1~cm based on air measurements. Note that the 1000 mm acrylic pipe doesn’t have any weight balances, thus the buoyancy effect does not need to be accounted for.


A short run of approximately one hour provides sufficient statistics (about 1,000,000 events) for a precise calibration at each position.   

\section{Event Selection for $^{252}$Cf Source Data}
\label{sec:method}
The $^{252}$Cf source emits prompt fission gammas simultaneously with neutrons. As the neutrons capture on the scintillator after a short delay, resulting in additional gamma emission, they are identified by detecting the coincidence of two signals within a short time window.
The selection criteria for this coincidence technique are detailed in Table~\ref{tab:Cfselection}. 
The total energy of the fission gammas can reach up to 20~MeV~\cite{CITE:Cffission}, thus an energy range of 12-20~MeV is required for prompt activities. A 12~MeV threshold is used to distinguish prompt activities from delayed activities.
\begin{table}
\begin{center}
\caption{\label{tab:Cfselection} The selection criteria for	neutron capture events from $^{252}$Cf.} 
\begin{tabular}{cccc}
 \\ \hline
 E (prompt) & E (Delayed) & timing & spatial corr.\\ 
 (MeV) & (MeV) & ($\mu$s) & (cm) \\  \hline
 12 - 20 & 7 - 12 & $\Delta t <$ 100 & $\Delta_{VTX} <$ 60\\ \hline
\end{tabular}
\end{center}
\end{table}
The calibration provides energy and vertex information, and an iterative process between the calibration and the event selection refines the samples for further studies.

\begin{figure}[h]
\begin{center}
\includegraphics[width=0.40\textwidth]{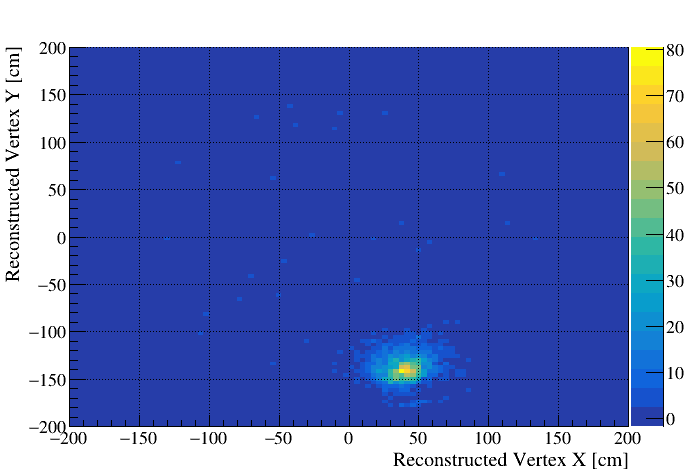} 
\caption{\label{fig:XY_R150} The reconstructed vertices of the n-Gd 
captured events on the x-y plane for the $R$ = 160.0~cm calibration.} 
\end{center}
\end{figure}

Figure~\ref{fig:XY_R150} displays the reconstructed vertices of n-Gd 
capture events in the x-y plane for the $R$ = 160.0~cm calibration, after applying the selection criteria. 

\section{Event Reconstruction Methodology}
\label{sec:JADE}
The event reconstruction algorithm in JSNS$^2$~\cite{CITE:JADE}
utilizes the observed charges of PhotoMultiplier Tube (PMT) signals 
and compares them with expected values based on an effective photocathode area calculation for a PMT. This calculation assumes the isotropic emission of scintillation light assumed to originate from a point source. The effective photocathode area is influenced by two main factors: the distance ($r$) between the point source and the PMT, following a 1/$r^2$ function, and the zenith angle ($\eta$) of the point source relative to the center of the PMT photocathode sphere.
Because the photocathode covers only a specific area of the PMT, the effective photocathode area also depends on the zenith angle.

The event reconstruction algorithm in JSNS$^2$ optimizes the vertex position and the total amount of scintillation light using charges recorded by 96~PMTs. The algorithm minimizes the difference between the observed and the expected charges of 96~PMTs, where the expected charges are calculated from $r$, the zenith angle, and the total amount of scintillation light. Each PMT's charge corresponds to a number of photoelectrons, calibrated by the LED system~\cite{CITE:NIM} within the detector.

Given the relatively compact size of the acrylic tank in the JSNS$^2$ detector, this approach yields good results, as demonstrated later.

\begin{figure}[!b]
\begin{center}
\includegraphics[width=0.48\textwidth]{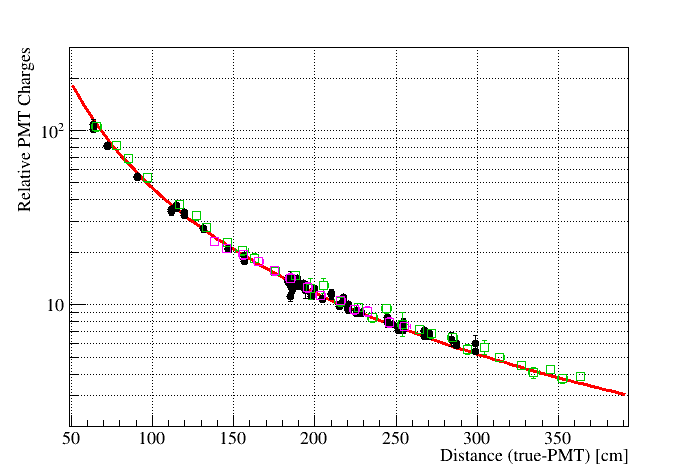} 
\includegraphics[width=0.48\textwidth]{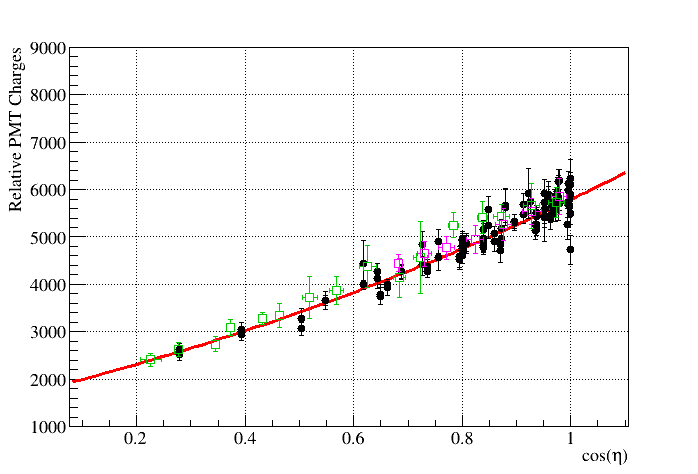} 
\caption{\label{fig:cos_func} The left panel shows the PMT charge as a function of $r$, while the right panel displays the relative PMT charge as a function of the zenith angle between the point source and the center of the photocathode sphere of the PMT. The black points represent the data from the one-dimensional calibration, while the magenta points (from the 1000~mm pipe) and the green points (from the 2200~mm pipe) correspond to the data from the three-dimensional system. The red lines represent the best fit to the data.} 
\end{center}
\end{figure}

Figure~\ref{fig:cos_func} depicts the observed relative PMT charge as a function of distance (left) and cosine of the zenith angle 
(right). In the left panel, the red line represents the $1/r^{2}$ function used in the JSNS$^2$ reconstruction, while the right panel 
shows the effective photocathode area as a function of the zenith angle, also depicted with a red line. The error bars in both panels include the systematic uncertainties arising from the PMT charge estimation. 

\begin{figure}[h]
\begin{center}
\includegraphics[width=0.47\textwidth]{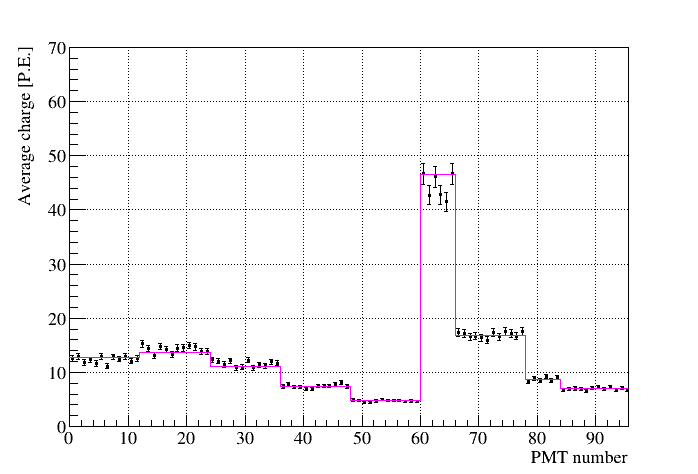} 
\caption{\label{fig:z75} The plot displays observed (black) and expected (magenta) PMT charges at the z = 75~cm position of the $^{252}$Cf source in one-dimensional calibration system. The horizontal axis represents the PMT number, structured as follows: Numbers 1-12 correspond to the top row of barrel PMTs out of 5 rows, totaling up to 60 (12 columns $\times$ 5 rows). Numbers 61-66 correspond to the inner PMT circle, while 67-78 
represent the outer circle of 12 top-lid PMTs. Numbers 79-96 represent bottom PMTs.} 
\end{center}
\end{figure}

For the effective photocathode area, no obvious physics models are available, so the data points are fitted using the following function~\eqref{eq:gx}.
\begin{equation}
    g(\xi) = D \Bigl[\frac{C}{1+e^{ (-A(\xi-B))}} + \bigl(1-\frac{C}{1+e^{ (-A(1-B))}} \Bigr) \Bigr],
    \xi = \cos(\eta)
    \label{eq:gx}
\end{equation}
where $A, B$, and $C$ are parameters for the slope horizontal axis shift of the exponential function and $D$ is a normalization factor.
The fit results for the parameters are summarized in Table~\ref{tab:fit}.
\begin{table}
\begin{center}
\caption{\label{tab:fit} The fit results of parameters of the zenith angle dependence of the charge. As described in the function~\eqref{eq:gx} in the main text, $A, B$, and $C$ are parameters for the slope horizontal axis shift of the exponential function and $D$ is a normalization factor.} 
\begin{tabular}{ccccc}
 \\ \hline
 A & B & C & D\\ \hline
 0.89$\pm$0.19 & 2.64$\pm$0.25 &  
 76.9703$\pm$1.01 & 
 5789$\pm$36 \\  \hline
\end{tabular}
\end{center}
\end{table}
The red lines demonstrate a reasonable agreement with the observed data. 
Figure~\ref{fig:z75} compares the observed and the expected 
PMT charges using the one-dimensional calibration data at coordinates 
(x,y,z) = (0,0,75~cm). Good agreement is observed between the two.

\section{Event Reconstruction Performance}
\label{sec:Cap}

\begin{figure}[h]
\begin{center}
\includegraphics[width=0.44\textwidth]{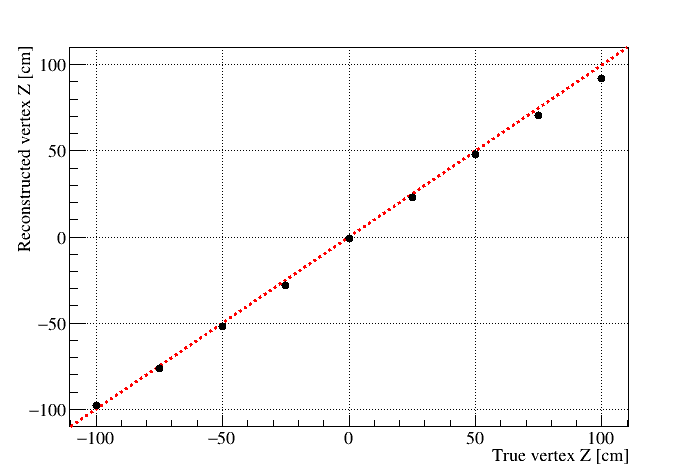} 
\includegraphics[width=0.44\textwidth]{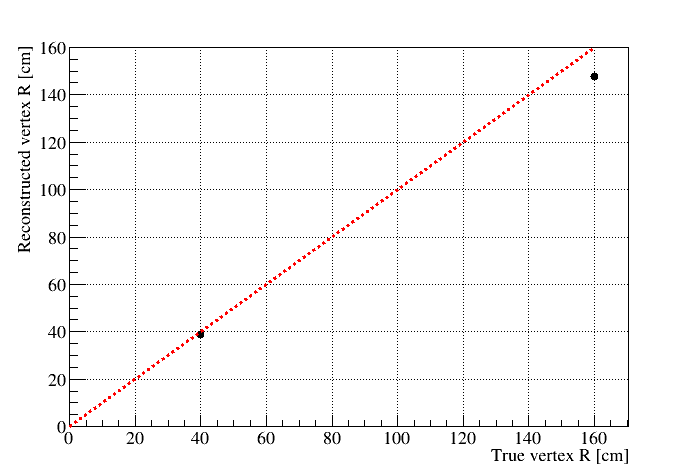} 
\caption{\label{fig:truerec} A comparison between the true and reconstructed vertex positions. The left panel shows the results from the one-dimensional calibration system, while the right panel shows results from the three-dimensional one.} 
\end{center}
\end{figure}

Figure~\ref{fig:truerec} illustrates the comparison between true and reconstructed vertices. The source positions of the calibration systems were measured in the air, providing the true vertices.
The left plot depicts results from the campaign with the one-dimensional calibration system, while the right plot shows the results of the campaign with the three-dimensional calibration system.
Within the JSNS$^2$ fiducial volume, the largest difference observed is 8~cm at Z = 100~cm. The fiducial volume is defined as the $R < 140$~cm and $|z| < 100$~cm region to minimize external backgrounds. The precision of the fiducial volume is determined to be 
5$\pm$0.7\% for the z axis and 7.6$\pm$0.7\% for $R$ using calibrations at the fiducial edge regions. The cylindrical shape of $R^2 - z$, yields an expected precision of 20\%  for the fiducial volume without additional corrections. It should be noted that the reconstructed values for both $R$- and $z$-directions systematically underestimate the true values.

\begin{figure}[h]
\begin{center}
\includegraphics[width=0.46\textwidth]{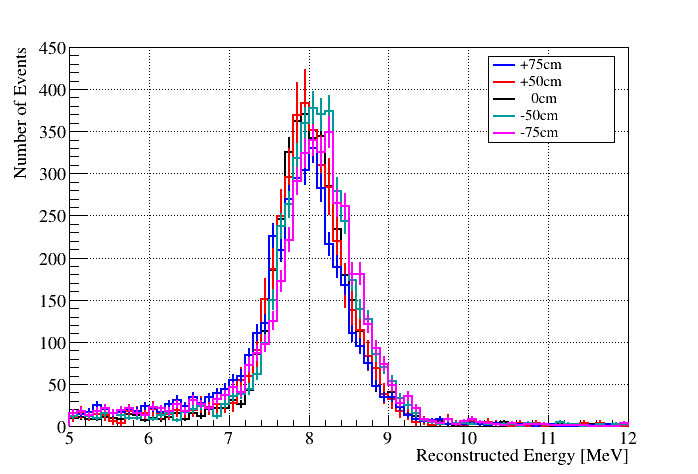} 
\caption{\label{fig:E8MeV} The reconstructed energy for the n-Gd  events at various calibration points of the one-dimensional calibration campaign.} 
\end{center}
\end{figure}

\begin{figure}[htb!]
\begin{center}
\includegraphics[width=0.8\textwidth]{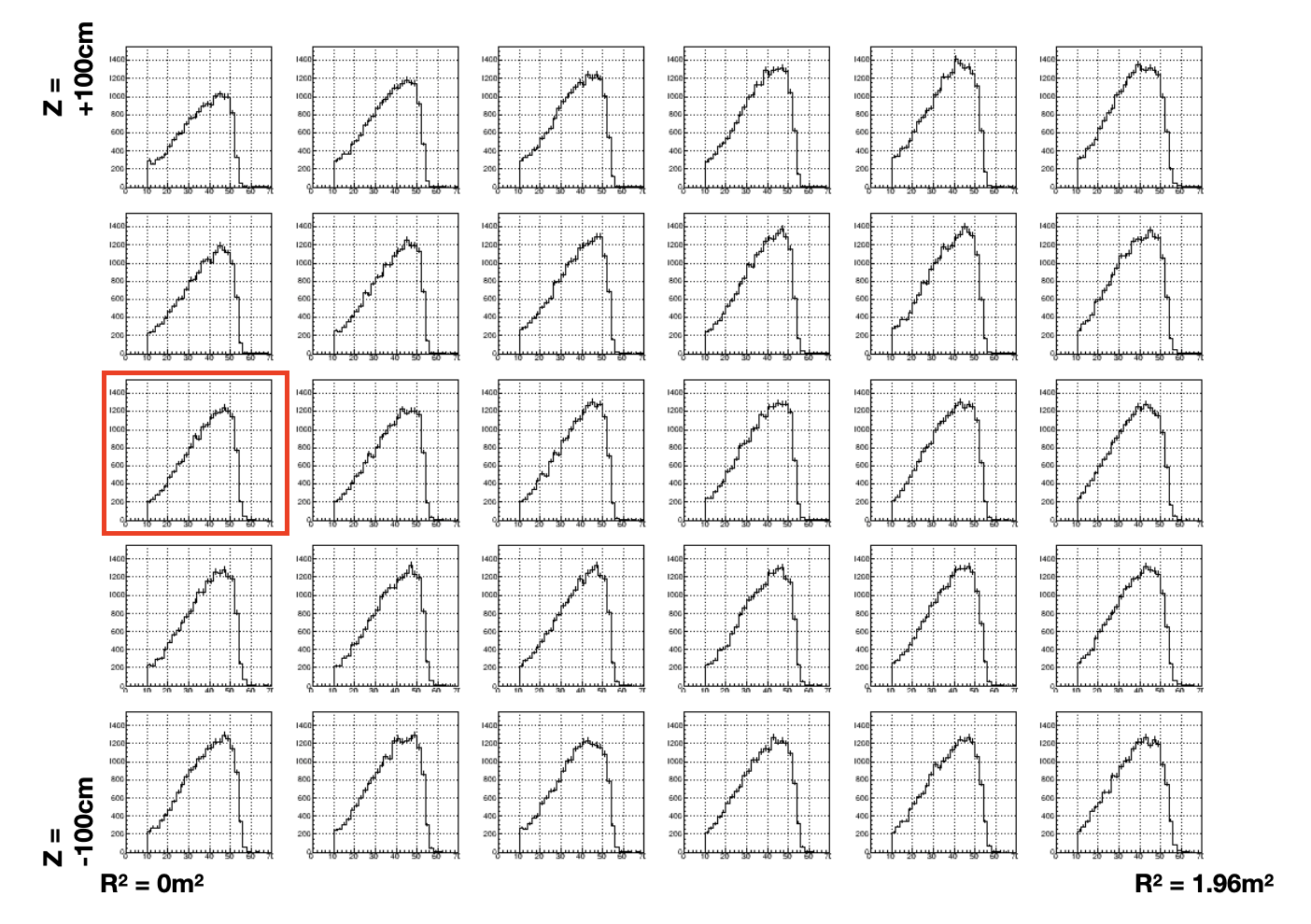}
\caption{\label{fig:EME} The reconstructed energy spectra for the Michel Electron sample. The vertical axis (z-axis) spans from -100~cm to +100~cm, and is divided into 5 intervals. The horizontal axis (R$^2$ axis) ranges from 0 to 1.96~m$^2$, and is divided into 6 intervals. The central pixel is highlighted with a red box.} 
\end{center}
\end{figure}

The reconstructed energy of neutron capture events is also shown in Fig.~\ref{fig:E8MeV}. 
In the five calibration points, the expected energy around 8~MeV is well reconstructed.
To assess the performance of the $^{252}$Cf calibration in reconstructing energies up to approximately 50~MeV for sterile neutrino searches, we studied cosmogenic Michel electron (ME) events. 
Figure~\ref{fig:EME} presents the energy spectra of the ME decays from the
stopped muons within the detector, organized by each pixel of our fiducial $R^2 - z$ plane. This illustrates the reconstruction capabilities of our analysis.

The selection criteria for the ME sample are shown in Table~\ref{tab:EventSelection}. To mitigate the cosmic muon interference, additional requirements include veto sums of less than 100~photo-electrons (p.e.) for the top and bottom 12 PMTS.
\begin{table}
\begin{center}
\caption{\label{tab:EventSelection} The selection criteria for the cosmogenic Michel electron sample.} 
\begin{tabular}{cccc}
 \\ \hline
 E (Prompt) & E (Delayed) & timing & spatial corr.\\ 
 (MeV) & (MeV) & ($\mu$s) & (cm) \\  \hline
 10 - 800 & 20 - 60 & $\Delta t < 10$ & $\Delta_{VTX} <$ 130\\ \hline
\end{tabular}
\end{center}
\end{table}
After selection, approximately 100~Hz of ME events are retained. The energy range of the ME sample is identical to that of the IBD signals, underscoring the importance of accurate reconstruction.
The energy spectra reconstructed across all fiducial regions demonstrate good performance.

\begin{figure}[htb!]
\begin{center}
\includegraphics[width=0.45\textwidth]{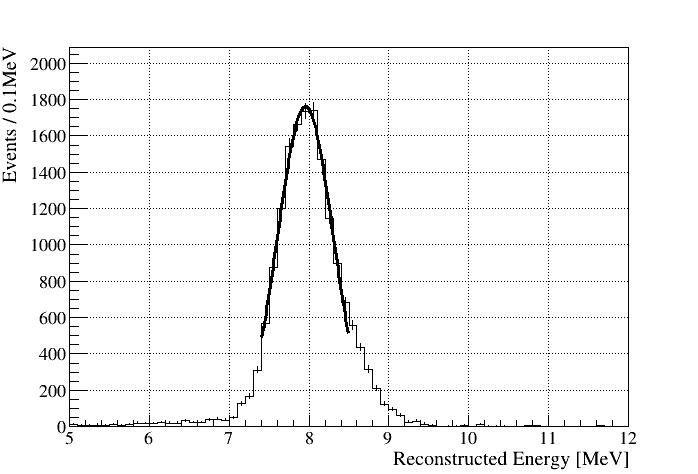}
\includegraphics[width=0.45\textwidth]{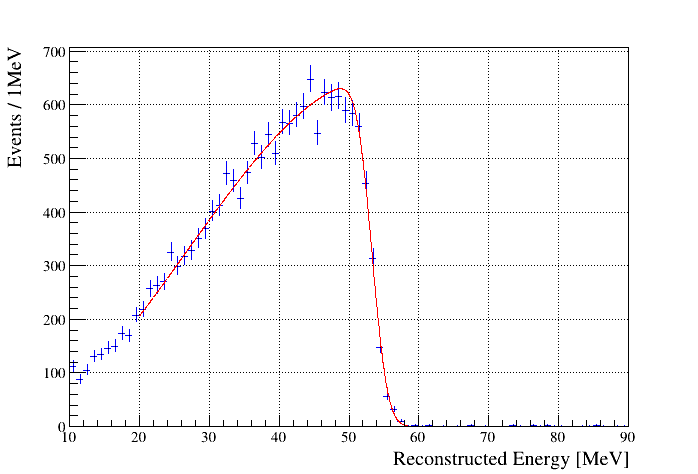} 
\caption{\label{fig:Eres} Fit results to evaluate the energy resolutions of the central region of the detector (z=0 cm for the n-Gd events (left) and the central pixel in Fig.~\ref{fig:EME} (right)).} 
\end{center}
\end{figure}

Energy resolution was evaluated using both n-Gd captured events and ME events. Figure~\ref{fig:Eres} displays the fitting results for assessing energy resolution in the central region of the detector (z=0 cm for n-Gd events and the central pixel in Figure~\ref{fig:EME}). A single Gaussian is applied to n-Gd events to calibrate the main peak, while the theoretical energy spectrum of Michel electrons, convoluted with a Gaussian, is used for ME events. 
The obtained resolutions are 4.3$\pm$0.1\% for the n-Gd events around  8~MeV, and 3.3$\pm$0.1\% for ME events around 53~MeV.

\section{Summary}
\label{sec:Summary}
We conducted a performance assessment of the JSNS$^2$ reconstruction algorithm (JADE) using a $^{252}$Cf radioactive source. The differences between the true and the reconstructed vertices
 are 5.0$\pm$0.7\% for the $z$-direction, and 7.6$\pm$0.7\% for the $R$-direction. Based on these results, the uncertainty of the fiducial volume of JSNS$^2$ is evaluated as 20\% without corrections. 
As mentioned earlier, the reconstructed vertices for both R and Z directions are systematically shift inside compared to the true vertices. Therefore, a 20\% correction is applied to the fiducial volume. After applying this correction, the uncertainty of the fiducial volume is much less than the requirement of JSNS$^2$ requirement (10\%). The dominant contribution of the uncertainty of the fiducial volume after this correction comes from the statistics of the $^{252}$Cf source calibration data if the relationship between the true calibration points and the reconstructed vertices are used.

To evaluate the energy-reconstruction capability, we utilized n-Gd capture events and cosmogenic Michel electrons. Both samples exhibit reasonable energy spectra, approximately 8~MeV for n-Gd events and approximately 53~MeV endpoint for Michel electrons. The energy resolution at the Michel endpoint is 3.3$\pm$0.1\% in the central 
region, while it is 4.3$\pm$0.1\% for n-Gd peak. 

\section{Acknowledgement}
\label{sec:Summary}
We extend our sincere gratitude to the J-PARC staff, particularly the MLF and accelerator groups, for their invaluable support in facilitating this experiment.

The work is supported by the Ministry of Education, Culture, Sports, Science and Technology (MEXT)
and the Japan Society for the Promotion of Science (JSPS) grants-in-aid: 16H06344, 16H03967, 23K13133, 23K13133 and 20H05624, Japan. 

Financial support from the National Research Foundation of Korea (NRF) is also gratefully acknowledged for the grants: 2016R1A5A1004684, 2017K1A3A7A09015973, 2017K1A3A7A09016426,
2019R1A2C3004955, 2016R1D1A3B02010606, 2017R1A2B4011200, 2018R1D1A1B07050425, 2020K1A3A7A09080133, 2020K1A3A7A09080114, 2020R1I1A3066835, 2021R1A2C1013661, 2022R1A5A1030700 and RS-2023-00212787. Additional support from a fund from the BK21 of the NRF is acknowledged. 

The University of Michigan gratefully acknowledges the support of the Heising-Simons Foundation. The work carried out at Brookhaven
National Laboratory was supported by the US Department of Energy under the DE-AC02-98CH10886 contract. The University of Sussex acknowledges support from the Royal Society grant no. IESnR3n170385. 

We are grateful to the Daya Bay Collaboration for providing the Gd-LS, the RENO Collaboration for providing the LS and PMTs, CIEMAT for providing the splitters, Drexel University for providing the FEE circuits, and Tokyo Institute of Technology for providing FADC boards.





\end{document}